\documentstyle[twocolumn,aps,epsf]{revtex}
\begin{document}
\newcommand{\beq}{\begin{equation}}
\newcommand{\eeq}{\end{equation}}
\title{AGING EFFECT IN CERAMIC SUPERCONDUCTORS}
\author{Mai Suan Li$^{1,2}$, Per Nordblad$^3$ and Hikaru Kawamura$^4$}
\address{$^1$Institute of Physics, Polish Academy of Sciences,
Al. Lotnikow 32/46, 02-668 Warsaw, Poland\\
$^2$ Institut f{\H u}r Theoretische Physik, Universit{\H a}t zu K{\H o}ln,
Z{\H u}lpicher Stra{\ss}e 77, D-50937 K{\H o}ln, Germany\\
$^3$ Department of Materials Science, Uppsala University,
Box 534, S-751 21 Uppsala, Sweden\\
$^4$ Department of Earth and Space Science, Faculty of Science,
Osaka University, Toyonaka, Osaka 560-0043, Japan}
\address{
\centering{
\medskip\em
{}~\\
\begin{minipage}{14cm}
A three-dimensional lattice
of the Josephson junctions with a finite self-conductance is employed
to model ceramic superconductors.
Using Monte Carlo simulations it is shown that the aging disappears
in the strong screening limit. In the weeak screening regime aging
is present even at low temperatures. For intermediate values of
the self-inductance aging occurs at intermediate temperatures
interval but is suppressed entirely at high and low
temperatures. Our results are in good agreement with experiments.
{}~\\
{}~
\end{minipage}
}}
\maketitle

The off-equilibrium dynamical properties of glassy systems have attracted
the attention of researchers for many years. In particular, aging phenomena
observed in spin glasses \cite{Lund} have
been studied both theoretically
\cite{Theory} and experimentally \cite{Exper} in detail.
In the aging phenomenon the physical quantities depend not only on
the observation time but also on the waiting time, $t_w$, i.e. how long one
waits at constant field and temperature before measurements.
The origin of such memory phenomena  relates to the rugged energy landscape
which appears due to disorder and frustration \cite{Binder}.

Recently, the aging effect has been observed in the ceramic superconductor
 Bi$_2$Sr$_2$CaCu$_2$O$_8$  by Papadopoulou {\em et al} \cite{Nordblad}
monitoring the zero field cooled (ZFC) magnetization. The relaxation of
the ZFC magnetization has been measured by cooling the sample in zero field
to the measuring temperature, allowing the sample to stay
at that temperature for a certain time $t_w$ and then applying the probing
field and recording the change of the magnetization with observation time
at constant temperature.
Papadopoulou {\em et al} have made two key observations. First,
the aging effect is not observed at high fields and at high temperatures.
Second, {\em at low temperatures this effect disappears again}.
Thus the aging phenomenon exists only for weak enough fields and
in an intermediate temperature region. The first
observation is trivial because at high fields or high temperatures the
roughness of the energy landscape does not play any crucial role and
the system looses its memory. Since at low temperatures the role of
the energy landscape becomes important,
the second result of Ref. \cite{Nordblad} is not trivial
from the point of view of the standard spin glass theory.
Papadopoulou {\em et al} suggested that at low temperatures the external
field is screened from the bulk of the sample and it cannot probe the
collective behavior of the Josephson junction network. One of the aims
of our paper is to check if this idea is correct.

It should be noted that in addition to the aging effect also
the paramagnetic Meissner effect (PME) \cite{Braunish}
was observed in the investigated Bi$_2$Sr$_2$CaCu$_2$O$_8$
compound \cite{Nordblad}. This result is important for the following reason.
It is well known that the PME in ceramic superconductors may be explained
based on the existence of $\pi$-junctions in
the Josephson network \cite{Sigrist}. Such $\pi$-junctions could arise
naturally as a consequence of $d$-wave pairing of the superconducting order
parameter. One of the potential candidates for this pairing is the
$d_{x^2-y^2}$ state. Despite the fact that the existence of $\pi$-junction
was confirmed experimentally by a scanning SQUID microscope \cite{Tsuei},
only the detection of the PME does not unambiguously
support the $d$-wave symmetry because PME has been
also observed in conventional superconductors \cite{Disk}. However,
the combined observation of aging and the PME in the same material does
yield support for the existence of $d$-wave superconductivity.

In order to explain the experiments \cite{Nordblad} we use a model
for the $d$-wave superconductivity in a Josepson junction network
\cite{Dominguez}. Using Monte Carlo simulations we demonstrate that
there are three screening regimes for the aging phenomenon. In the
strong screening limit the aging is suppressed at any temperature.
In the weak screening regime it is observable even at low
temperatures. The intermediate screening regime is  found
to be the most interesting: the aging is here present only in an
intermediate temperature interval, it does not appear at
low temperatures. This is precisely what was observed in the
experiments \cite{Nordblad}.
The underlying mechanism of aging is twofold: the screening
makes the energy landscape less rough and the external magnetic field
is screened from the bulk. The latter effect is strong at low temperatures
and at high values of the inductance.

We neglect the charging effects of the grain and
consider the following Hamiltonian\cite{Dominguez,KawLi}
\begin{eqnarray}
{\cal H} = - \sum _{<ij>} J_{ij}\cos (\theta _i-\theta _j-A_{ij})+ \nonumber\\
\frac {1}{2L} \sum _p (\Phi_p - \Phi_p^{ext})^2, \nonumber\\
\Phi_p \; \; = \; \; \frac{\phi_0}{2\pi} \sum_{<ij>}^{p} A_{ij} \; , \;
A_{ij} \; = \; \frac{2\pi}{\phi_0} \int_{i}^{j} \, \vec{A}(\vec{r})
\cdot
d\vec{r} \; \; ,
\end{eqnarray}
where $\theta _i$ is the phase of the condensate of the grain
at the $i$-th site of a simple cubic lattice,
$\vec A$ is the fluctuating gauge potential at each link
of the lattice,
$\phi _0$ denotes the flux quantum,
$J_{ij}$ denotes the Josephson coupling
between the $i$-th and $j$-th grains,
$L$ is the self-inductance of a loop (an elementary plaquette),
while the mutual inductance between different loops
is neglected.
The first sum is taken over all nearest-neighbor pairs and the
second sum is taken over all elementary plaquettes on the lattice.
Fluctuating  variables to be summed over are the phase variables,
$\theta _i$, at each site and the gauge variables, $A_{ij}$, at each
link. $\Phi_p$ is the total magnetic flux threading through the
$p$-th plaquette, whereas $\Phi_p^{ext}$ is the flux due to an
external magnetic field applied along the $z$-direction,
\begin{equation}
\Phi_p^{ext
} = \left\{ \begin{array}{ll}
                   HS \; \;  & \mbox{if $p$ is on the $<xy>$ plane}\\
                   0  & \mbox{otherwise} \; \; ,
                        \end{array}
                  \right.
\end{equation}
where $H$ and $S$ denote the external magnetic field and
the area of an elementary plaquette, respectively.
In what follows we assume $J_{ij}$ to be an independent random variable
taking the values $J$ or $-J$ with equal probability ($\pm J$ or bimodal
distribution), each representing 0 and $\pi$ junctions.

\begin{figure}
\epsfxsize=3.2in
\centerline{\epsffile{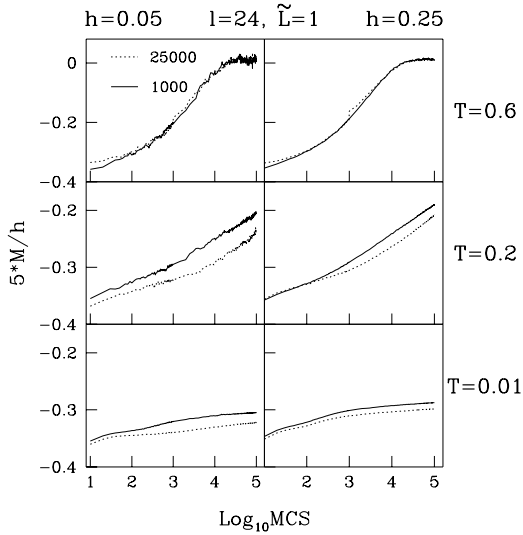}}
\caption{The temperature and time dependence of $M$
for $t_w=1000$ (solid line) and $t_w=25000$ (dotted line).
$l=24, \tilde L=1$ , $h=0.25$ and $h=0.05$.
The results are averaged over
60 -- 120 samples.}
\end{figure}

It should be noted that the model (1) captures not only
the PME \cite{Dominguez,KawLi} but also several dynamical phenomena of
ceramic high-$T_c$ superconductors, such as:
the anomalous microwave absorption \cite{Dominguez1},
the so called compensation effect \cite{Li}, the effect
of applied electric fields in the apparent critical current \cite{DWJ}
and the AC resistivity \cite{Li1}. Similar to the spin glass case,
the frustration due to the random
distribution of $\pi$-junctions should lead to
a multivalley energy landscape with energy barriers separating different
metastable states. Such a rugged energy landscape would favor
the aging effect in the model (1). In fact, an extensive of Monte Carlo
simulation by two of the authors  \cite{Kawa1}
has revealed that the model (1)
in zero field exhibits an equilibrium phase transition with
a broken time-reversal symmetry into the novel ''chiral glass'' state.
In this chiral glass phase, ''chiralities'', or local loop supercurrents
circulating over grains carrying a half flux quantum, are
frozen in a spatially random manner.

The dimensionless magnetization along the $z$-axis normalized per plaquette,
$M$, is given by
\begin{equation}
M \; \; = \; \; \frac{1}{N_p\phi_0} \; \sum_{p\in <xy>}
(\Phi_p - \Phi_p^{ext}) \; \; ,
\end{equation}
where the sum is taken over all $N_p$ plaquettes on the $<xy>$ plane of the
lattice.
The dimensionless field
$h$ and inductance
$\tilde L$ are defined
as follows
\begin{eqnarray}
h \; \; = \; \; \frac{2\pi H S}{\phi_0} \; \; , \; \;
\tilde L \; \; = \; \; (2\pi /\phi_0)^2 JL.
\end{eqnarray}
The parameter $\tilde L$ controls different screening regimes:
the larger the $\tilde L$, the stronger the screening.

\begin{figure}
\epsfxsize=3.2in
\vspace{0.1in}
\centerline{\epsffile{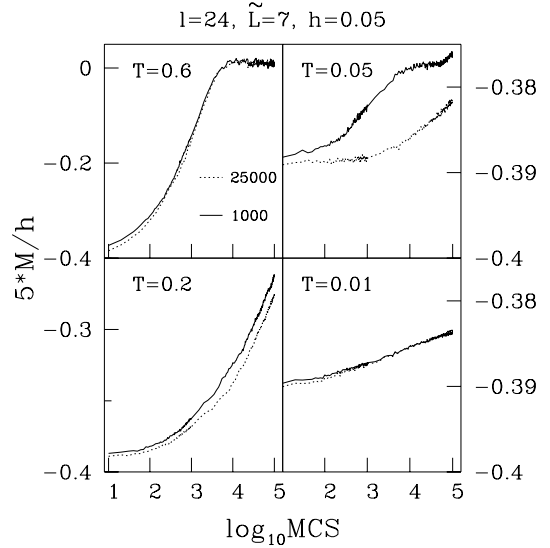}}
\vspace{0.2in}
\caption{The temperature and time dependence of $M$ for
$t_w=1000$ and $t_w=25000$.
$l=24, \tilde L=7$ and $h=0.05$. The results are averaged over
60 -- 120 samples.}
\end{figure}

In order to study the aging effect in the zero field cooled (ZFC)
regime we quench the system from a high temperature to
the working temperature. The system is there evolved in zero field during
a waiting time, $t_w$. Then the external field $h$ is turned on and
the subsequent growth of the magnetization $M(t,t_w)$ is monitored.
The free boundary conditions are implemented
(the magnetization always vanishes for the periodic boundary conditions
\cite{Dominguez,KawLi}). 

We have checked  the finite size effect for three-dimensional systems 
of linear sizes $l=12, 24$ and 36.
Since this effect is not substantial
we will present the results for $l=24$.
Following experiments \cite{Nordblad}, we choose
observation times to be of order of $t_w$.
Fig. 1 shows the dependence of the magnetization on $t$ and $t_w$
for $l=24$, $\tilde L=1$, $h=0.25$ and 0.05 for several
typical temperatures (measured in units of $J$).
Note that for this inductance the chiral glass transition takes
place at $T\simeq 0.286$ \cite{Kawa1}.
The time is measured
in the number of Monte Carlo steps (MCS).
For $h=0.05$ (left panel) the aging effect disappears at high temperatures.
The screening effect is visible at low $T$'s but it is not strong
enough to suppress the aging effect entirely.
One may think that the screening would become more important for stronger
fields and the aging would not occur at low temperatures. Our results
for $h=0.25$ (right panel) show, however, that this is not the case.
In what follows we will focus on $h=0.05$, because a twice as small value
of $h$ does not change the results in any substantial way.
We have found that for $\tilde L < \tilde L_1^*$
(weak screening regime), where
the borderline value $\tilde L_1^*=3.5 \pm 0.5$, one
has the standard scenario: the aging
disappears only at high temperatures.

\begin{figure}
\epsfxsize=3.2in
\vspace{0.1in}
\centerline{\epsffile{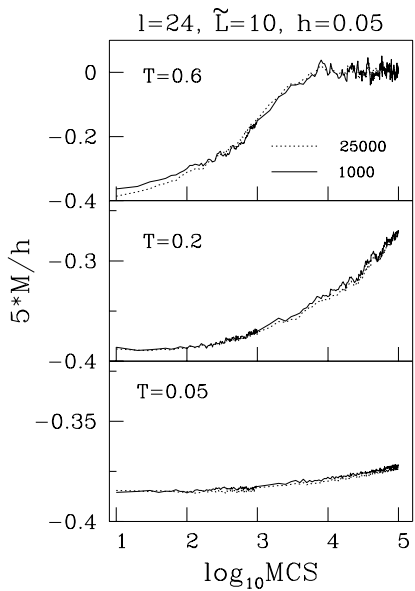}}
\vspace{0.2in}
\caption{The temperature and time dependence of $M$ for
$t_w=1000$ and $t_w=25000$.
$l=24, \tilde L=10$ and $h=0.05$. The results are averaged over
120 -- 240 samples.}
\end{figure}

Fig. 2 shows the results for $\tilde L$=7.
In agreement with the experiments \cite{Nordblad},
the aging effect appears only for the intermediate temperature interval.
At low $T$'s the effect is suppressed due to the screening of the magnetic
field from the bulk. This is the main result of the present paper.
We have found that the non-standard scenario is observed for
$\tilde L_1^* < \tilde L < \tilde L_2^*$, where
the borderline value $\tilde L_2^*=9.0 \pm 0.5$.
For this interval of $\tilde L$ the aging effect disappears at temperatures
$T \le T^* = 0.02 \pm 0.01$.
Our results have been obtained for the observation times comparable with the
waiting ones but we believe that they should be valid for longer observation
time scales.
It should be stressed that the experimental finding of
Papadopoulou {\em et al} \cite{Nordblad} cannot be explained by the
standard XY model where the screening effect is not taken into account.

One can demonstrate that in the strong screening limit
$\tilde L > \tilde L_2^*$, the aging effect does not occur
at any temperature because the external field is screened entirely.
It may be seen in Fig. 3 where the results for $\tilde L$=10
are presented.
The present model exhibits a finite-temperature chiral glass
ordering for
$\tilde L < \tilde L_{c}^{CG} $ (the superscript means
the chiral glass),
where $\tilde L_{c}^{CG}=6 \pm 1$ \cite{Kawa1}.
Since $\tilde L_2^* > \tilde L_{c}^{CG}$
the aging effect is suppressed in the region where the chiral
glass phase is not favored. The PME is, however, observed for any strength
of screening \cite{KawLi}.
The observation of both the PME and the aging phenomenon supports the
hypothesis about the existence of $\pi$-junctions (and
$d$-wave pairing) because the two effects cannot occur simultaneously in
the flux compression picture for the PME \cite{Larkin}.

\begin{figure}
\epsfxsize=3.2in
\vspace{0.1in}
\centerline{\epsffile{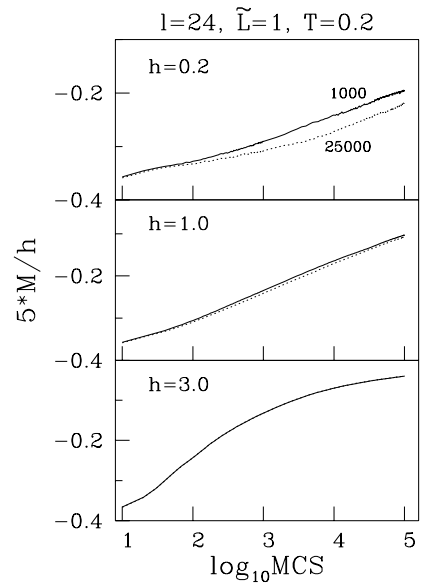}}
\vspace{0.2in}
\caption{The field and time dependence of $M$ for $t_w=1000$ and $t_w=25000$.
$l=24, \tilde L=1$ and $T=0.2$. The results are averaged over
20 -- 60 samples.}
\end{figure}
\noindent

Fig. 4 shows the results for $T=0.2$ and different values of $h$.
In agreement with the experiments \cite{Nordblad}, the aging effect gets more
and more suppressed as the field increases.
For sufficiently large magnetic fields, $h > 1$, this effect is no longer
observable.

The question we ask now is what is the mechanism of aging behaviour
in our model. We have studied the spatial distribution of flux, 
$\tilde{\Phi}(r)$,
inside the sample (the definition of $\tilde{\Phi}(r)$ is
given in Ref. \cite{KawLi}). Fig. 5 shows such distribution obtained as an averaged
$\tilde{\Phi}(r)$ over the observation time and over four
equivalent directions along $\pm x$ and $\pm y$ axes. 
Here $r$ is a distance from the 
surface in units of lattice spacing.
For a fixed screening strength the magnetic field is expelled more
and more from the bulk as $T$ is lowered. Therefore, for the strong
enough screening one could not observe the aging at low $T$'s.
The results in the lower panel of Fig. 5 show that the stronger screening
the weaker penetration of the field into the sample. Consequently, the aging
effect should be suppressed in the strong screening limit.
Three distinct scenarios of the aging phenomenon may be understood
qualitatively based on the flux distribution inside the sample.

We have also explored the effect of screening on the 
energy landscape.
Our preliminary studies of local minima at $T=0$ show that the energy landscape
gets smoother as the screening is increased \cite{Li2} and
the glassy effects would become less pronounced.

\begin{figure}
\epsfxsize=3.2in
\vspace{0.1in}
\centerline{\epsffile{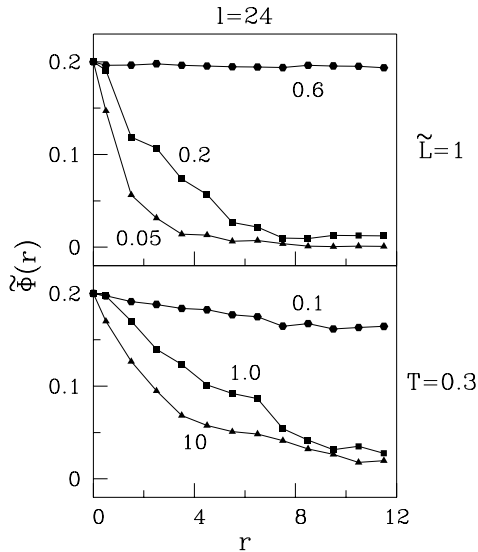}}
\vspace{0.2in}
\caption{Spatial flux distributions for several values of $T$
and $\tilde{L}$.
The upper panel corresponds to $\tilde{L}=1$ and $T=0.6, 0.2$ and 0.05.
The values of $T$ are shown next to the curves.
The lower panel corresponds to $T=0.3$ and $\tilde{L}=0.1, 1$ and 10.
The values of $\tilde{L}$ are shown next to the curves.
We take $l=24$ and $h=0.2$. Depending on $T$ and
$\tilde{L}$ the results are averaged over
30 - 50  samples.}
\end{figure}

In conclusion, our study reveals that screening has a strong
influence on the aging phenomenon. For
$\tilde L_1^* < \tilde L < \tilde L_2^*$, aging
appears only at intermediately high temperatures. This non-trivial behaviour
agrees with recent experimental results for
a ceramic superconductor.
The study of the aging effect sheds new light
on the important problem about the nature of the symmetry of the
superconducting order parameter. On the other hand, the aging effect was
also found to have
a strong correlation with the
occurence of the chiral glass phase. Experimental search for this phase
would be of great interest.

\noindent

Financial support from the Polish agency KBN
(Grant number 2P03B-025-13 and Grant number 2P03B-146 18) is acknowledged.
MSL thanks T. Nattermann for the warm hospitality at the Institute for
Theoretical Physics, Cologne University.
\par


\noindent
\begin{references}
\bibitem{Lund} L. Lundgren, P. Svedlindh, P. Nordblad,
and O. Beckman,
Phys. Rev. Lett. {\bf 51}, 911 (1983).
\bibitem{Theory} J. -P. Bouchaud, L. F. Cugliandolo, J. Kurchan, and M. Mezard
in {\em Spin Glasses and Random Fields}
(World Scientific, Singapore, 1998), ed. A. P. Young, p. 161
\bibitem{Exper} P. Nordblad and P. Svedlindh, {\em Ibid}, p. 1.
\bibitem{Binder} K. Binder and A. P. Young, Rev. Mod. Phys. {\bf 58},
801 (1986).
\bibitem{Nordblad} E. L. Papadopoulou, P. Nordblad, Svendlindh,
S. Sch{\H o}neberger, and R. Gross , Phys. Rev. Lett.
{\bf 82}, 173 (1999).
\bibitem{Braunish} P. Svedlindh,  K. Niskanen, P. Nordblad, L. Lundgren,
B. L\"onnberg and T. Lundstr\"om, Physica C {\bf 162-164}, 1365 (1989);
 W. Braunish,  N. Knauf, V. Kataev, S. Neuhausen, A. Grutz, A. Kock,
B. Roden, D. Khomskii, and D. Wohlleben,
Phys. Rev. Lett. {\bf 68}, 1908 (1992).
\bibitem{Sigrist} M. Sigrist and T.M. Rice,
J. Phys. Soc. Jpn. {\bf 61}, 4283 (1992); Rev. Mod. Phys.
{\bf 67}, 503 (1995).
\bibitem{Tsuei} C. Tsuei, J. R. Kirtley, C. C. Chi, Lock See Yu-Jahnes,
A. Gupta, T. Shaw, J. Z. Sun, and M. B. Ketchen,  Phys. Rev. Lett.
{\bf 73}, 593 (1994);
J. Mannhart, H. Hilgenkamp, B. Mayer, Ch. Gerbe,
J. R. Kirtley, K. A. Moler, and M. Sigrist, {\em Ibid} {\bf 77},
2782 (1996).
\bibitem{Disk}
D. J. Thompson,
M. S. M. Minhaj, L. E. Wegner, and J. T. Chen,
Phys. Rev. Lett. {\bf 75}, 529 (1995);
P. Kostic,
B. Veal, A. P. Paulikas, U. Welp,
V. R. Todt, C. Gu, U. Geiser, J. M. Williams, K. D. Carlson, and
R. A. Klemm,
Phys. Rev. B {\bf 53}, 791 (1996); {\bf 55}, 14649 (1997);
L. Pust, L. E. Wegner, and M. R. Koblischka,
Phys. Rev. B {\bf 58}, 14191 (1998);
A. K. Geim ,
S. V. Dubonos, J. G. S. Lok, M. Henini,
and J. C. Maan,
Nature {\bf 396}, 144 (1998).
\bibitem{Dominguez} D. Dom\'inguez, E.A. Jagla and C.A. Balseiro,
Phys. Rev. Lett. {\bf 72}, 2773 (1994).
\bibitem{KawLi} H. Kawamura and M.S. Li, Phys. Rev. B
{\bf 54}, 619 (1996).
\bibitem{Dominguez1} D. Dom\'{\i}nguez, E. A. Jagla,
and C. A. Balseiro, Physica
(Amsterdam) {\bf 235C - 240C}, 3283 (1994).
\bibitem{Li} M. S. Li, Phys. Rev. B  {\bf 60}, 118 (1999).
\bibitem{DWJ} D. Dom\'{\i}nguez, C. Wiecko and J. V. Jos\'{e},
Phys. Rev. Lett.  {\bf 83}, 4164 (1999).
\bibitem{Li1} M. S. Li and D. Dominguez, Phys. Rev. B. {\bf 62},
xxx (2000).
\bibitem{Kawa1} H. Kawamura and M. S. Li, Phys. Rev. Lett.
{\bf 78}, 1556 (1997);
H. Kawamura and M. S. Li,
J. Phys. Soc. Jpn {\bf 66}, 2110 (1997).
\bibitem{Larkin} A. E. Koshelev and A. I. Larkin, Phys. Rev. B {\bf 52},
13559 (1995); A. E. Khalil, Phys. Rev. B {\bf 55}, 6625 (1997),
V. V. Moshchalkov, X. G. Qui, and V. Bruyndoncz,
Phys. Rev. B {\bf 55}, 11793 (1997).
\bibitem{Li2} M. S. Li, unpublished.
\end{references}
\end{document}